\documentclass[pre,a4paper,floatfix,twocolumn]{revtex4}
\usepackage{epsfig,amsmath}
\begin{document}
\title{Noncollapsing solution below $r_c$ for a randomly forced
particle} \author{L. Anton} \email{anton@ifin.nipne.ro}
\affiliation{Institute for Theoretical Physics, University of
Stellenbosch, Private Bag X1, 7602 Matieland, South Africa}
\affiliation{Institute of Atomic Physics, INFLPR, Laboratory 22, P.O.
Box MG-36 R76900, Bucharest, Romania} \date{\today}

\begin{abstract}
We show that a non collapsing solution bellow $r_c$ can be constructed
for the dynamics of randomly forced particle interacting with a
dissipating boundary. The scaling analysis predicts a divergent
collision rate at the boundary for the non-collapsing solution. This
prediction is tested numerically.
\end{abstract}

\maketitle
%&latex209
%\documentstyle[aps,epsfig]{revtex}
%\begin{document}
%\draft
%\wideabs{
%\title{Non-collapsing solution bellow $r_c$ for randomly forced particle} 
%\author{L. Anton\cite{emailaddress}}
%\address{Institute for Theoretical Physics, University of
%Stellenbosch, Private Bag X1, 7602 Matieland, South Africa \\ and\\
%Institute of Atomic Physics, INFLPR, Lab 22, PO Box MG-36 R76900,
%Bucharest, Romania} \date{\today} \maketitle

%}%end wide abs

Recently opposing viewpoints appeared in the literature
\cite{Cornell98,Cornell00,Florencio00} regarding the localization
properties of a one-dimensional particle subject to an uncorrelated
random force interacting with a dissipating boundary.  The
equation describing the particle dynamics is
\begin{equation}\label{Langevininit}
\frac{d^2x}{dt^2}=\eta(t)
\end{equation}
where $\eta(t)$ is Gaussian white noise, $\langle \eta(t)\eta(t')\rangle=
2\delta (t-t')$.  The dissipating boundary condition is set such that
the particle approaching with velocity $-u$, $(u>0)$, at the boundary
$x=0$ is reflected with velocity $ru$, $r<1$.  The problem was
explored further with various techniques
\cite{Smedt00,Burkhardt00b,Burkhardt01}, which have obtained the same
critical value for the dissipation parameter $r_c$ and the persistence
exponent. It is an unresolved problem, the absence of the
collapsing behavior in numerical simulations \cite{Florencio00}.

In this paper, we propose a solution of this paradox. We show that
for $r<r_c$ one can construct a constant mass (noncollapsing)
solution starting from the collapsing one. In our approach, we use the
Fokker-Planck equation (FPE) description of the process Eq.
(\ref{Langevininit}). The FPE associated with the Langevin equation
(\ref{Langevininit}) is
\begin{equation}\label{FP}
\biggl(\frac{\partial}{\partial t}-\frac{\partial^2}{\partial u^2}+
u \frac{\partial}{\partial x}\biggr)P(x,u,t)=0
\end{equation}
with the dissipating boundary condition 
\begin{equation}\label{bc}
P(0,-u)=r^2P(0,ru),\quad u>0,
\end{equation}
and the initial condition $P(x,u,t=0)=\delta(x-x_0)\delta(u-u_0)$.

In Ref. \cite{Burkhardt01}, it was shown that the the general
solution of this problem  has the following integral form
\begin{equation}\label{gensol}
\begin{array}{ll}
&P(x,u;x_0,u_0;t)=P_0(x,u;x_0,u_0;t)\\
&+\int_{0}^{t}dt_1\int_{0}^{\infty}du_1 u_1 P(x,u;0,ru_1;t-t_1)
P_0(0,-u_1;x_0,u_0;t_1),
\end{array}
\end{equation}
where $P_0(x,u;x_0,u_0;t)$ is the solution of the FPE Eq. (\ref{FP})
with absorbing boundary at $x=0$ \cite{Burkhardt93}.

\begin{figure}
%\includegraphics[bb= 50pt 30pt 550pt 750pt]{fig1.ps}
%\caption{\label{fig1} The first return distribution for
%$r=0.1,\;0.05,\;0$ ($r_c\approx 0.163$). The lines have the
%theoretical exponent found in \protect\cite{Burkhardt00b}.}
%\begin{minipage}[t]{0.9\linewidth}
\epsfig{file=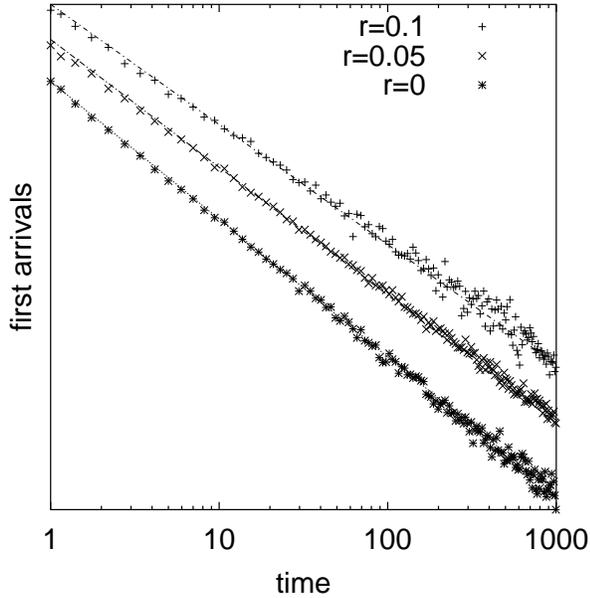,clip=,%
%height=0.20\textheight,%
width=0.45\textwidth, bbllx=50pt,bblly=30pt,bburx=550pt,bbury=550pt}
\caption{\label{fig1}First return distribution for
$r=0.1,\;0.05,\;0$ ($r_c\approx 0.163$). The lines have the
theoretical exponent found in Ref. \protect\cite{Burkhardt00b}.}
%\end{minipage}
\end{figure}

Burkhardt has shown in Ref. \cite{Burkhardt00b} that a collapsing
solution with an algebraic temporal decay can be found for the
Eq. (\ref{FP}). The surviving probability $Q(x_0,u_0,t)=\int dxdv
P(x,u;x_0,u_0,t)$ behaves asymptotically as
\begin{eqnarray}
Q(0,-u,t)&\approx & 2\sin\Bigl[\frac{\pi}{6}(1-4\phi)\Bigr]
\Bigl(\frac{u^2}{t}\Bigr)^\phi, 
\quad u>0\label{q1}\\ 
Q(0,u,t)&\approx & \Bigl(\frac{u^2}{t}\Bigr)^\phi,
\quad u>0.\label{q2}
\end{eqnarray}
For the collapsing solution ($\phi >0$) we have $Q(0,0,t)=0$, that is,
the origin of the phase space $(x=0,u=0)$ is an absorbing point for
the random particle. With this observation the collapsing behavior can
also be obtained numerically. Using a discretization of
Eq. (\ref{Langevininit}) together with the absorbing prescription at
the origin, the collapsing behavior is obtained with the persistence
exponent in perfect agreement with theoretical result of Ref.
\cite{Burkhardt00b}, see Fig.~\ref{fig1}. In numerical simulation the
particle was absorbed if its velocity after the collision with the
boundary was smaller than $\sqrt{\Delta t}$, where $\Delta t$ is the
time integration step. As $\Delta t$ decreases the collapsing behavior
is preserved signaling the existence of the collapsing behavior in
the continuum limit.

On the other hand, Eqs. (\ref{q1}), (\ref{q2}) accept the solution
$\phi=0$ at any value of the dissipation coefficient $r$ case in which
the collapse does not occur since $Q$ is constant.
%
%Nevertheless if the absorbing condition is not imposed the collapsing
%behavior does not appear in the numerical simulation, fact reported
%first time in \ref{}.
%
We make the second observation that Eq. (\ref{gensol}) is not in
contradiction with a solution that has constant mass, that is,
$\int_{0}^{\infty} dx \int_{-\infty}^{\infty} du
P(x,u;x_0,u_0;t)=1$. Indeed if one integrates the over $x$ and $u$ the
right hand side of Eq. (\ref{gensol}) is the conservation law for the
particle in case of the absorbing solution. The reason that
allows the existence of more than one solution for this problem is
that the dissipating condition $P(0,-u)=r^2P(0,ru)$ does not
specify uniquely the solution on the boundary $x=0$, but is just a
condition that the solution must obey on the boundary $x=0$. The
continuity condition for the solution asks that $P(0,u\rightarrow
0)\rightarrow 0$ or $\infty$. 

We can construct a noncollapsing solution starting from the
collapsing one. Let us start from the FPE with a source term
\begin{equation}
\biggl(\frac{\partial}{\partial t}-\frac{\partial^2}{\partial u^2}+ u
\frac{\partial}{\partial
x}\biggr)P(x,u,t)=f(t)\delta(x-\epsilon)\delta(u),
\end{equation}
where $\epsilon >0$ and $f(t)$ is at the moment an arbitrary function
of time to be determined from the conserving mass condition.
The solution of the above equation that satisfies the boundary
condition Eq. (\ref{bc}) is

\begin{equation}
\begin{array}{l}
G_r(x,u;x_0,u_0;t)=P_r(x,u;x_0,u_0;t)\\
+\int_{0}^{t}dt_1 P_r(x,u;\epsilon,0;t-t_1)f(t_1),
\end{array}
\end{equation}
where $P_r(x,u;x_0,u_0;t)$ is the collapsing solution of the problem
at given $r<r_c$.
After time Laplace transform, we have
\begin{equation}\label{sol}
G(x,u;x_0,u_0;s)=P_r(x,u;x_0,u_0;s)+
P_r(x,u;\epsilon,0;s)f(s).
\end{equation}
We can choose  $f(s)$ such that the mass of $G$ is constant. We
see that the choice
\begin{equation}
f(s)=\frac{\frac{1}{s}-\int_{0}^{\infty}dx_1\int_{0}^{\infty} du_1
P_r(x_1,u_1;x_0,u_0;s)} {\int_{0}^{\infty}dx_1\int_{-\infty}^{\infty}
du_1 P_r(x_1,u_1;\epsilon,0;s)}
\end{equation}
gives us the needed solution. For any $\epsilon >0$ the solution
Eq. (\ref{sol}) satisfies the boundary condition since we use $P_r$,
and conserves the mass by construction. In the limit
$\epsilon\rightarrow 0$ it satisfies also the initial condition
$\delta(t)\delta(x-x_0)\delta(u-u_0)$. We make the observation that
$f(s)=0$ for $r>r_c$ as $P_r(x,u;x_0,u_0;t)$ has a constant mass. It
remains to be shown that the limit $\epsilon\rightarrow 0$ exists. For
the case $r=0$ the asymptotic expressions for $P_r(x,u;x_0,u_0;s)$ and
$\int_{0}^{\infty}\int_{-\infty}^{\infty} du P_r(x,u;x_0,u_0;s)$ were
obtained in Ref. \cite{Burkhardt93} and one can see explicitly that
the above limit exists.

We can obtain the behavior of the collision rate at the origin for
small $\epsilon$ using the scaling properties of the solution of the
FPE.  Equation (\ref{FP}) gives
\begin{equation}\label{scaling}
P(x,u;x_0,u_0;s)=\lambda^{2} P(\lambda^3 x,\lambda
u;\lambda^3 x_0,\lambda u_0;\lambda^{-2}s).
\end{equation}
The collision rate $R_{\text{coll}}$ is given by the following
relations:
\begin{align*}
& R_{\text{coll}}(\epsilon;s)=\int_{-\infty}^{0}du\; u
G(0,u;x_0,u_0;s)\\&=\int_{-\infty}^{0}du\; u
P_r(0,u;x_0,u_0;s)\\  & +\frac{\int_{-\infty}^{0}du\; u
P_r(0,u;\epsilon,0;s)} {\int_{0}^{\infty}
dx\int_{0}^{\infty}du\;
P_r(x,u;\epsilon,0;s)}\\&
\times\Bigl[\frac{1}{s}-\int_{0}^{\infty}
\!dx\int_{0}^{\infty}\!du\; P_r(x,u;x_0,u_0;s)\Bigr].
\end{align*}

Using the scaling property, Eq. (\ref{scaling}), with
$\lambda=\epsilon^{-1/3}$ we have for the term depending on $\epsilon$:
\begin{eqnarray}\label{bounces}
&&\frac{J(\epsilon;s)}{Q(\epsilon; s)}=\frac{\int_{-\infty}^{0}du\; u
P_r(0,u;\epsilon,0;s)} {\int_{0}^{\infty}
dx\int_{0}^{\infty}du\;
P_r(x,u;\epsilon,0;s)}\nonumber\\
&&=\frac{\tilde J(\epsilon^{2/3}
s)}{\epsilon^{2/3}\tilde Q(\epsilon^{2/3} s)}\approx
\epsilon^{-2\phi/3},\;\;\epsilon \ll 1,
\end{eqnarray}
where we have used that $\tilde J(s)\approx 1-c s^{\phi}$ and $\tilde
Q(s)\approx s^{-1+\phi}$ for $s\ll 1$.

Equation (\ref{bounces}) shows that the conserving solution has
divergent collision rate at the origin. This implies that the the
current density $uG(0,u;x_0,y_0;s)$ is nonintegrable, as was first
noted in Ref.  \cite{Burkhardt00a}.

\begin{figure}[t]
%\includegraphics[bb= 50pt 30pt 550pt 750pt]{fig2.ps}
%\begin{minipage}[t]{0.9\linewidth}
\epsfig{file=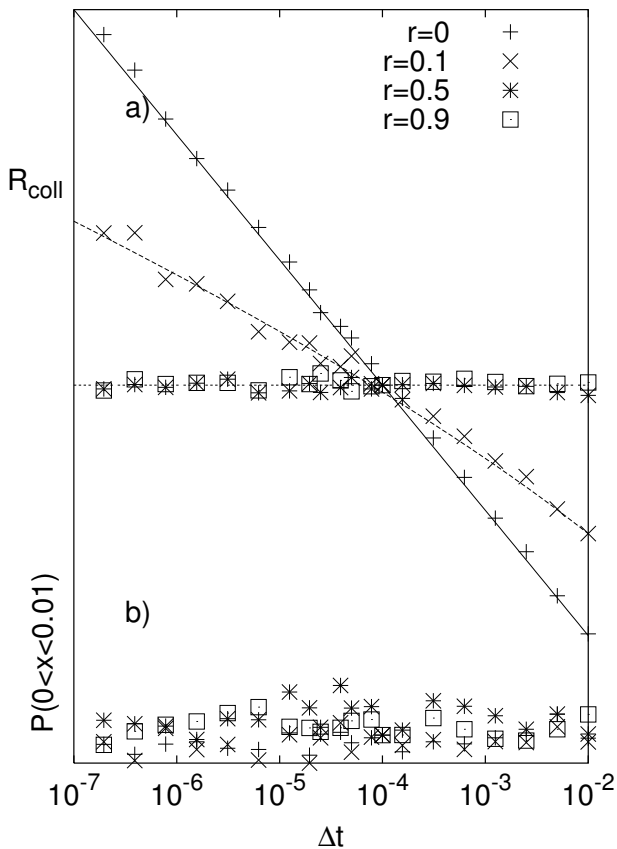,%
width=0.45\textwidth,%
bbllx=50pt,bblly=30pt,bburx=550pt,bbury=750pt}
\caption{\label{fig2}a) Collision rate at $x=0$ function of
integration time step $\Delta t$ at various values of the restitution
coefficient $r$. We see that for $r>r_c\approx 0.163$ the collision
rate is independent of $\Delta t$ whereas diverges like $(\Delta
t)^{-\phi}$ for $r<r_c$. The lines plot the theoretical prediction
with $\phi$ $=0.25 (r=0),\approx 0.087(r=0.1), =0(r\ge r_c)$. For
$r=0.1$, we considered also the subleading correction. b) Probability
to find the particle in the interval $(0,0.01)$ function of $\Delta
t$. The quantity is constant for both $r<r_c$ and $r>r_c$. The graphs
were displaced vertically for clarity.}
%\end{minipage}
\end{figure}

The prediction of a divergent collision rate can be checked
numerically. If we integrate the Langevin equation with a finite time
step $\Delta t$, then the particle is injected at $\epsilon\approx
(\Delta t)^{3/2}$, where $\epsilon$ is its velocity that is very
small. Consequently, the number of bounces at the origin must diverge
as $(\Delta t)^{-\phi}$ as $\Delta t\rightarrow 0$. Indeed
Fig.~\ref{fig2} shows a perfect validation of this prediction. In the
same figure we have plotted the probability that the particle stays
between $x=0,\;x=0.01$. We see that this this probability is
independent of the integration step for each value of the restitution
parameter $r$.  This means that there is no collapsing behavior. The
particle is attracted by the wall, performs an infinite number of
collisions (in the limit $\Delta t\rightarrow 0$) and with probability
$1$ is injected back into the domain $x>0$.

Now we can see that the question ''what the particle does if we put it
at $x=0$, $u=0$?' is indeterminate for $r<r_c$. The general solution
in this case is
\begin{equation}
\begin{array}{r}
P=qG_r(x,u,x_0,u_0,t)+(1-q)P_r(x,u,x_0,u_0,t),\\
 0\le q\le 1,
\end{array}
\end{equation} 
and one has to specify $q$ for the answer.

In conclusion, we have shown that the collapsing behavior can be found
numerically if one notice that the collapsing solution has an
absorbing point into the origin, hence it must be enforced in the
simulation.

We have constructed a noncollapsing solution for the case
$r<r_c$. This is possible since the absorbing boundary condition
$P(0,-u)=r^2P(0,ru)$ allows for two functions as boundary
condition. One goes to zero as $u\rightarrow 0$ and the other one
diverges to $\infty$ as $u\rightarrow 0$.

In terms of Brownian paths, we propose the following picture: for
$r>r_c$ the probability to touch the origin of the phase space
starting from any other point is zero, similar to simple diffusion in
two or more dimensions, thus the collapsing solution is forbidden. When
$r<r_c$ the diffusing particle touches the origin with probability $1$
and if the path is set to terminate there the collapse occur. If the
path is not set to terminate the particle is sent back into the domain
$x>0$ after an infinite number of collision with the boundary. The
weight of the paths leaving the origin without touching the boundary is
zero in the continuum limit but they give a finite contribution
because they are sampled an infinite number of times.
  
The author thanks Ted Burkhardt, Alan Bray, Michael Swift, Stephen
Cornell, H.B. Geyer, F.G. Scholtz, and A. van Biljion for
criticism and useful discussions. The final version of the paper was
prepared at Department of Physics, Inha University, South Korea.

%\bibliographystyle{apsrev}
%\bibliography{/home/anton/papers/biblio}

\end{document}